\documentclass[intlimits,twoside,a4paper]{article}
\usepackage{amsmath,amssymb}
\usepackage{graphicx}
\usepackage{dcolumn}% Align table columns on decimal point
\usepackage{threeparttable}
\usepackage[T2A]{fontenc}
\usepackage[cp1251]{inputenc}

\newcolumntype{d}[1]{D{.}{.}{#1}}% or D{.}{,}{#1} or D{.}{\cdot}{#1}

\usepackage{cmpj2}
\issue{2012}{15}{3}{33701}
\doinumber{10.5488/CMP.15.33701}

\title[Investigations of the $g$ factors for fresh PrBa$_{2}$Cu$_{3}$O$_{6+x}$ powders]%
{Investigations of the $g$ factors and local structure for orthorhombic $\text{Cu}^{2+}(1)$ site in
fresh $\text{PrBa}_{2}\text{Cu}_{3}\text{O}_{6+x}$ powders}
\author[M.Q. Kuang \textsl{et al.}]{M.Q. Kuang\thanks{E-mail: {mqkuang@yeah.net}} \,\,\refaddr{label1},
S.Y. Wu\refaddr{label1,label2},
B.T. Song\refaddr{label1},
X.F. Hu\refaddr{label1}}

\authorcopyright{M.Q. Kuang, S.Y. Wu, B.T. Song, X.F. Hu, 2012}

\addresses{
\addr{label1}Department of Applied Physics, University of
Electronic Science and Technology of China, \\Chengdu 610054,
P.R.China
\addr{label2}International Centre for Materials Physics,
Chinese Academy of Science, \\Shenyang 110016, P.R.China
}

\date{Received November 28, 2011, in final form February 16, 2012}

\begin{document}

\maketitle

\begin{abstract}

The electron paramagnetic resonance (EPR) $g$ factors $g_x$,
$g_{y}$ and $g_{z}$ of the orthorhombic $\text{Cu}^{2+}(1)$ site in
fresh $\text{PrBa}_{2}\text{Cu}_{3}\text{O}_{6+x}$ powders are
theoretically investigated using the perturbation formulas of the $g$ factors for a $3d^{9}$ ion under orthorhombically elongated
octahedra. The local orthorhombic distortion around the
$\text{Cu}^{2+}(1)$ site due to the Jahn-Teller effect is described
by the orthorhombic field parameters from the superposition model.
The $\text{[CuO6]}^{10-}$ complex is found to experience an axial
elongation of about 0.04~{\AA} along $c$ axis and the relative bond
length variation of about 0.09~{\AA} along $a$ and $b$ axes of the
Jahn-Teller nature. The theoretical results of the $g$ factors based
on the above local structure are in reasonable agreement with the experimental data.
\keywords crystal fields and Hamiltonians, electron paramagnetic resonance, $\text{Cu}^{2+}$,
$\text{PrBa}_{2}\text{Cu}_{3}\text{O}_{6+x}$
\pacs 71.70.Ch, 74.25.Nf, 74.72.Bk
\end{abstract}

\section{Introduction}

$\text{PrBa}_{2}\text{Cu}_{3}\text{O}_{6+x}$ (Pr123) compounds
are useful materials with anomalous resistive and magnetic  \cite{1},
vortex  \cite{2,3}, friction  \cite{4}, structural  \cite{5} and
superconductive  \cite{6} properties and have attracted extensive
interest of researchers. These properties are largely related to the
local structure and electronic behaviours near the $\text{Cu}^{2+}$
site, which can be investigated by means of the electron
paramagnetic resonance (EPR) technique. For example, EPR experiments
were carried out for fresh
$\text{PrBa}_2\text{Cu}_{3}\text{O}_{6+x}$ powders, and the
anisotropic $g$ factors $g_{x}$, $g_{y}$ and $g_{z}$ were also
measured for the $\text{Cu}^{2+}(1)$ site  \cite{7}. Until now,
however, the above experimental results have not been quantitatively
interpreted, and the local structure around the $\text{Cu}^{2+}(1)$
site is not determined, either. Since the electronic properties and
the local structure of the paramagnetic $\text{Cu}^{2+}$ as well as
the microscopic mechanisms of its EPR spectra would be helpful in
understanding the properties of the Pr123 systems, further theoretical
investigations on the $g$ factors and the local structure of the
$\text{Cu}^{2+}(1)$ site for the fresh
$\text{PrBa}_{2}\text{Cu}_{3}\text{O}_{6+x}$ powders are of
fundamental and practical significance. In this work, the $g$ factors
and the local structure of the fresh
$\text{PrBa}_{2}\text{Cu}_{3}\text{O}_{6+x}$ powders are
theoretically studied using the high order perturbation formulas of
the $g$ factors for a $3d^{9}$ ion in an orthorhombically elongated
octahedron. In the calculations, the local orthorhombic distortion
of the $\text{Cu}^{2+}(1)$ site is quantitatively involved from the
superposition model in view of the Jahn-Teller effect.

\section{Calculation}

 In the orthorhombic phase of
$\text{PrBa}_{2}\text{Cu}_{3}\text{O}_{6+x}$, the
$\text{Cu}^{2+}(1)$ site belonging to orthorhombic point symmetry
$(D_{2})$ has six nearest neighbour oxygen ligands, which construct
a distorted octahedron with the approximately mutually vertical Cu--O
bonds of the average (or reference) distance $R$ (${}\approx  1.917$~{\AA}  \cite{8}). As a Jahn-Teller ion, $\text{Cu}^{2+}$ can suffer
the Jahn-Teller effect via relaxation and compression of the Cu--O
bonds parallel with and perpendicular to the $c$ axis in terms of the
relative axial elongation $\Delta Z$. Meanwhile, the planar Cu--O
bonds may suffer another relative bond length variation $\Delta X$
along $a$ and $b$ axes. Thus, the local structure of the orthorhombic
$\text{Cu}^{2+}(1)$ site in the fresh
$\text{PrBa}_{2}\text{Cu}_{3}\text{O}_{6+x}$ powders can be
characterized by axial elongation $\Delta Z$ and the
perpendicular bond length variation $\Delta X$ (see figure~1).

 For a $\text{Cu}^{2+}$($3d^{9}$) ion in an orthorhombically
elongated octahedron, the original cubic ground orbital doublet
$^{2}E_{g}$ may split into two orbital singlets $^{2}A_{1g}$ and
$^{2}A_{1g}'$, with the latter lying lowest. Nevertheless, the upper
cubic orbital triplet $^{2}T_{2g}$ can be separated into three
orbital singlets $^{2}B_{1g}$, $^{2}B_{2g}$ and $^{2}B_{3g}$
 \cite{9}. The high order perturbation formulas of the $g$ factors for
an orthorhombically elongated octahedron can be expressed as follows
 \cite{10}:
\begin{eqnarray}
g_{x}&=&g_{\mathrm{s}}+2\emph{k}\frac{\zeta}{E_{2}}+\emph{k}\zeta^{2}\left[\frac{2}{E_{1}E_{2}}
     -\frac{1}{E_{2}E_{3}}-\frac{4}{E_{1}E_{3}}\right]
     +g_{\mathrm{s}}\zeta^{2}\left[\frac{2}{E_{1}^{2}}-\frac{1}{2E_{2}^{2}}+\frac{1}{2E_{3}^{2}}\right]\nonumber\\
     & &{}-\emph{k}\zeta^{3}\left[\frac{\left(\frac{1}{E_{2}}-\frac{1}{E_{3}}\right)\left(\frac{1}{E_{3}}+\frac{1}{E_{2}}\right)}{2E_{1}}
     + \frac{\left(\frac{2}{E_{1}}-\frac{1}{E_{2}}\right)\left(\frac{2}{E_{1}}+\frac{1}{E_{2}}\right)}{2E_{3}}
     - \frac{\frac{1}{E_{2}}-\frac{1}{E_{3}}}{2E_{2}E_{4}}\right]\nonumber\\
     & &{}+\frac{g_{\mathrm{s}}\zeta^{3}}{4}\left[\frac{\frac{1}{E_{3}}-\frac{2}{E_{1}}}{E_{2}^{2}}
     +\frac{\frac{2}{E_{3}}-\frac{1}{E_{2}}}{E_3^{2}}
     +2\frac{\frac{1}{E_{2}}-\frac{1}{E_{3}}}{E_{1}^{2}}
     +2\frac{\frac{1}{E_{2}^{2}}-\frac{1}{E_3^{2}}}{E_{1}}\right],\nonumber\\
g_{y}&=&g_{\mathrm{s}}+ 2\emph{k}\frac{\zeta}{E_{3}}+
\emph{k}\zeta^{2}\left[\frac{\frac{2}{E_{1}}-\frac{1}{E_{2}}}{E_{3}}-\frac{4}{E_{1}E_{2}}\right]
+g_s\zeta^{2}\left[\frac{2}{E_{1}^{2}}-\frac{1}{2E_{3}^{2}}+\frac{1}{2E_{2}^{2}}\right]\nonumber\\
     & &{}+\emph{k}\zeta^{3}\left[\frac{\left(\frac{1}{E_{2}}-\frac{1}{E_{3}}\right)\left(\frac{1}{E_{3}}+\frac{1}{E_{2}}\right)}{2E_{1}}
     +\frac{\left(\frac{2}{E_{1}}-\frac{1}{E_{3}}\right)\left(\frac{2}{E_{1}}+\frac{1}{E_{3}}\right)}{2E_{2}}
     -\frac{\frac{1}{E_{3}}-\frac{1}{E_{2}}}{2E_{3}E_{4}}\right]\nonumber\\
     & &{}+\frac{g_{\mathrm{s}}\zeta^{3}}{4}\left[\frac{\frac{1}{E_{2}}-\frac{2}{E_{1}}}{E_{3}^{2}}
     +\frac{\frac{2}{E_{2}}-\frac{1}{E_{3}}}{E_{2}^{2}}
     +2\frac{\frac{1}{E_{3}}-\frac{1}{E_{2}}}{E_{1}^{2}}
     +2\frac{\frac{1}{E_{3}^{2}}-\frac{1}{E_{2}^{2}}}{E_{1}}\right],\nonumber\\
g_{z}&=&g_{\mathrm{s}}+8\emph{k}\frac{\zeta}{E_{1}}
     +\emph{k}\zeta^{2}\left[\frac{1}{E_{2}E_{3}}+2\left(\frac{1}{E_{1}E_{2}}+\frac{1}{E_{1}E_{3}}\right)\right]
     -g_{\mathrm{s}}\zeta^{2}\left[\frac{1}{E_{1}^{2}}-\frac{\frac{1}{E_{2}^{2}}+\frac{1}{E_{3}^{2}}}{4}\right]\nonumber\\
     & &{}+\emph{k}\zeta^{3}\frac{\frac{8}{E_{1}}-\frac{1}{E_{2}}-\frac{1}{E_{3}}}{2E_{2}E_{3}}
     - 2\emph{k}\zeta^{3}\frac{\frac{1}{E_{1}E_{2}}+\frac{1}{E_{1}E_{3}}-\frac{1}{E_{2}E_{3}}}{E_{1}}\nonumber\\
     & &{}+\frac{g_{\mathrm{s}}\zeta^{3}}{4}\left[2\frac{\frac{1}{E_{2}^{2}}+\frac{1}{E_{3}^{2}}}{E_{1}}
     -\frac{\frac{1}{E_{2}}+\frac{1}{E_{3}}}{E_{2}E_{3}}\right].
     \label{eq1}
\end{eqnarray}
\begin{figure}[ht]
\centerline{\includegraphics[width=7cm]{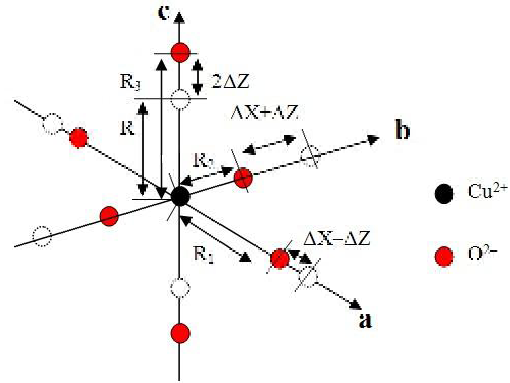}}
\caption{
The local structure of the $\text{Cu}^{2+}$(1) site in
the fresh $\text{PrBa}_{2}\text{Cu}_{3}\text{O}_{6+x}$ powders. The
local Cu--O bond lengths $R_{i}$ are described as the reference
distance $R$ in terms of the relative axial elongation $\Delta Z$ and
the planar bond length variation $\Delta X$.}
\end{figure}

 Here $g_{\mathrm{s}}$(${}\approx  2.0023$) is the spin-only value. $k$ is the orbital reduction factor, and  $\zeta$ is the spin-orbit
coupling coefficient of the $3d^{9}$ ion in crystals. The
denominators $E_{i}$ ($i = 1-3$) are the separations between the
excited $^{2}B_{1g}$, $^{2}B_{2g}$ and $^{2}B_{3g}$ and the ground
$^{2}A_{1g}'$ states, which can be obtained in terms of the cubic
field parameter $D_q$ and the orthorhombic field parameters $D_s$,
$D_t$, $D_{ \xi}$ and $D_{\eta}$ as  \cite{10}:
\begin{eqnarray}
E_{1}&=&10D_q,\nonumber\\
E_{2}&=&10D_q-3D_s+5D_t+3D_{\xi}-4D_{\eta}\,,\nonumber\\
E_{3}&=&10D_q-3D_s+5D_t-3D_{\xi}+4D_{\eta}\,.
\end{eqnarray}

 As mentioned before, the orthorhombic distortion of the
$\text{Cu}^{2+}(1)$ site may be described as the reference distance
$R$ as well as the relative axial elongation
 $\Delta Z$ and the planar bond length variation $\Delta X$. Thus, the Cu--O bond lengths along $a$, $b$ and $c$  axes are expressed as:  $R_{1} = R - \Delta Z + \Delta X$, $R_{2} = R - \Delta Z - \Delta X$ and
  $R_{3} = R+ 2\Delta Z$, respectively (see figure~1). This means that the Cu--O bonds
  suffer a relative elongation and compression of $2\Delta Z$ and $\Delta Z$ along $c$ and $a$ (or $b$) axes, respectively.
  Meanwhile, the planar Cu--O bonds undergo an additional relative bond length
variation of $\Delta X$ along $a$ and $b$ axes.

 Usually, the crystal-field parameters can be determined from
the superposition model which has been extensively adopted for
transition-metal ions in crystals  \cite{11}. Moreover, this model
can also be used for powder or polycrystal systems (e.g., the
superposition model analysis of the EPR spectra for $\text{Fe}^{3+}$
modified polycrystalline $\text{PbTiO}_{3}$  \cite{12} and
$\text{PbZrO}_{3}$  \cite{13} and $\text{Eu}^{2+}$ in polycrystalline
A zeolite  \cite{14}) and may be suitably applied to the fresh
$\text{PrBa}_{2}\text{Cu}_{3}\text{O}_{6+x}$ powders studied here. From the
local geometry and the superposition model  \cite{11}, the
orthorhombic field parameters can be determined as follows:
\begin{eqnarray}
D_{\mathrm{s}}&\approx&-2\overline{A}_{2}\left[(\emph{R}/\emph{R}_{1})^{t2}+(\emph{R}/\emph{R}_{2})^{t2}-2(\emph{R}/\emph{R}_{3})^{t2}\right]\big/7,\nonumber\\
D_{\xi}&\approx&2
\overline{A}_{2}\left[(\emph{R}/\emph{R}_{1})^{t2}-(\emph{R}/\emph{R}_{2})^{t2}\right]\big/7,\nonumber\\
D_{t}&\approx&8\overline{A}_{4}\left[2(\emph{R}/\emph{R}_{3})^{t4}-(\emph{R}/\emph{R}_{1})^{t4}-(\emph{R}/\emph{R}_{2})^{t4}\right]\big/21,\nonumber\\
D_{\eta}&\approx&5\overline{A}_{4}\left[(\emph{R}/\emph{R}_{1})^{t4}-(\emph{R}/\emph{R}_{2})^{t4}\right]\big/21\,.
\end{eqnarray}

  It is noted that the angular dependence is reduced due to the
 Cu--O bond angles (0, $\pi/2$ and
$\pi$ related to $Z$ axis and 0, $\pi/2$, $\pi$ and $3\pi/2$ related
to $X$ axis) and thus only the bond lengths $R_{i}$ are reserved in
the above formulas. Here $t_{2} \approx   3$ and $t_{4} \approx
5$ are the power-law exponents  \cite{11} and they are the intrinsic
parameters. For $3d^{n}$ ions in octahedra, the relationships
$\overline{A}_{4}\approx  3D_q/4$ and $\overline{A}_{2}\approx 10.8\overline{A}_{4}$  \cite{11,15,16} were proved to be valid for
many crystals and are suitably adopted here. Thus, the $g$ factors
(especially the axial anisotropy $\Delta g = g_{z}- (g_{x} +
g_{y})/2$ and the perpendicular anisotropy $\delta g = g_{x}-
g_{y}$) are correlated with the orthorhombic field parameters and
hence with the local structure of the system studied.

 From the optical spectra for Cu$^{2+}$ in some
oxides  \cite{17}, the spectral parameters $D_q {}\approx  1400 $~cm$^{-1}$
and $k\approx 0.77$ are obtained for the $\text{Cu}^{2+}$
center in the fresh $\text{PrBa}_{2}\text{Cu}_{3}\text{O}_{6+x}$
powders here. The spin-orbit coupling coefficient is usually
expressed as $\zeta \approx k \zeta_{0}$ , where
$\zeta_{0} (\approx 829 $~cm$^{-1}$  \cite{18}) is the corresponding
free-ion value. Substituting these values into equation~\eqref{eq1} and fitting
the calculated $g$ factors (especially the anisotropies) to the
experimental data, one can determine the local axial elongation and
the planar bond length variation:
\begin{eqnarray}
 \Delta Z  \approx 0.04~\text{{\AA}} \quad       \text{and} \quad         \Delta X  \approx   0.09 ~\text{{\AA}}  .
\end{eqnarray}
 The corresponding theoretical $g$ factors are shown in table~1.

\begin{table}
\caption{The anisotropic $g$ factors for the $\text{Cu}^{2+}(1)$ site
in the fresh $\text{PrBa}_{2}\text{Cu}_{3}\text{O}_{6+x}$ powders.}
\begin{center}  \begin{threeparttable}
   \begin{tabular}{|cc|c|c|c|c|c|c|c|c|}
   \hline
     & \ & $ g_{x}$ & $g_{y}$ & $g_{z}$ & $\Delta g$ & $\delta g$ \\
     \hline
     \hline
     & $Cal.$   & $2.056$    & $2.088$    & $2.224$    & $0.152$ & $0.032$\\
     \hline
     & {Expt}. \cite{7} & $2.050(4)$ & $2.094(4)$ & $2.222(4)$ & $0.15(8)$ & $0.044(8)$\\
     \hline
   \end{tabular}
  \end{threeparttable}
\end{center} \end{table}

\section{Discussion}

Table~1 reveals that the theoretical $g$ factors and the
anisotropies for the fresh
$\text{PrBa}_{2}\text{Cu}_{3}\text{O}_{6+x}$ powders based on the
local structural parameters $\Delta Z$ and $\Delta X$ are in good
agreement with the experimental data. Therefore, the EPR spectra are
satisfactorily interpreted for the fresh
$\text{PrBa}_{2}\text{Cu}_{3}\text{O}_{6+x}$ powders in this work.

 The EPR $g$ factors of the fresh
$\text{PrBa}_{2}\text{Cu}_{3}\text{O}_{6+x}$ powders can be
characterized by the axial and perpendicular anisotropies $\Delta g$
(${}\approx  0.15$) and $\delta g$ (${}\approx 0.044$), which are
ascribed to the local axial elongation $\Delta Z$ ($\approx
0.04$~{\AA}) and the planar bond length variation $\Delta X$ (${}\approx 0.09$~{\AA}), respectively. Thus, the $\text{Cu}^{2+}(1)$
site exhibits a moderate orthorhombic distortion of the Jahn-Teller
nature. Similar axial elongations and planar bond length variations
due to the Jahn-Teller effect were also found for some Jahn-Teller
ions (e.g., $\text{Cr}^{5+}$ and $\text{Ti}^{3+}$ with the same spin
$S=1/2$) in oxygen octahedra  \cite{19,20}. It seems that
$\text{Cu}^{2+}$ prefers to exhibit elongation distortions (i.e.,
orthorhombically elongated octahedra) under oxygen environments.

 There are some errors in the above calculations. First, the
approximations of the theoretical model and formulas may bring
about some errors in this work. Second, the errors also arise from
the approximation of the relationship $\overline{A}_{2}(R)$
${}\approx  10.8 \overline{A}_{4}(R)$  \cite{11,15,16}, which would
somewhat affect the orthorhombic field parameters and the final
results. The errors for the local structural parameters $\Delta Z$
and $\Delta X$ are estimated to be no more than 1\% as the ratio
$\overline{A}_{2}(R)/ \overline{A}_{4}(R)$ changes by 10\%. Third,
the present calculations are based on the conventional crystal-field
model containing only the central ion orbital and spin-orbit
coupling contributions, while the ligand orbital and spin-orbit
coupling contributions
 are not taken into account. Fortunately, although the studied system has
 some covalency (characterized by the covalency factor $N  \approx 0.77 < 1$), the spin-orbit
 coupling coefficient (${}\approx  151 $~cm$^{-1}$  \cite{21}) of the ligand $\text{O}^{2-}$ is much smaller than
 that (${}\approx  829 $~cm$^{-1}$  \cite{18}) of $\text{Cu}^{2+}$. According to various EPR studies for
 $\text{Cu}^{2+}$ under oxygen octahedra  \cite{22,23,24}, the ligand contributions to the $g$ factors may be very small and negligible. So, the present
theoretical calculations can be regarded as reasonable.
Moreover, the investigations in this work would be helpful in carrying out structural
  and magnetic studies on $\text{PrBa}_{2}\text{Cu}_{3}\text{O}_{6+x}$ superconductors
 as well as applicable to other similar R123 systems.

\section*{Acknowledgements}
 This work was supported by ``the Fundamental Research Funds
for the Central Universities''.

\newpage

\ukrainianpart

\title{Дослідження  $g$ факторів і локальної структури орторомбічного
вузла в чистому порошку $\text{PrBa}_{2}\text{Cu}_{3}\text{O}_{6+x}$}
\author{М.К. Куанг\refaddr{label1}, С.Й. Ву\refaddr{label1,label2}, Б.Т. Сонг\refaddr{label1}, Х.Ф. Ху\refaddr{label1}}

\addresses{
\addr{label1}Кафедра прикладної фiзики, Китайський унiверситет
електронiки та технологiй, Ченду 610054, КНР
\addr{label2}Мiжнародний центр фiзики матерiалiв, Академiя наук
Китаю, Шеньян 110016, КНР }

\makeukrtitle

\begin{abstract}
\tolerance=3000%
$g$ фактори  $g_x$, $g_{y}$ і $g_{z}$ електронного парамагнітного
резонансу для орторомбічного вузла  $\text{Cu}^{2+}(1)$ в чистому
порошку  $\text{PrBa}_{2}\text{Cu}_{3}\text{O}_{6+x}$ теоретично
досліджуються, використовуючи формалізм теорії збурень для цих
параметрів для $3d^{9}$ іона в орторомбічному видовженому октаедріi.
Локальна орторомбічна дисторція навколо вузла  $\text{Cu}^{2+}(1)$,
спричинена ефектом Яна-Теллера, описується за допомогою параметрів
орторомбічного поля із суперпозиційної моделі. Знайдено, що комплекс
$\text{[CuO6]}^{10-}$ піддається аксіальному видовженню близько
0.04~{\AA} вздовж  $c$ осі і відносна зміна довжини зв'язку природи
Яна-Теллера є приблизно 0.09~{\AA} вздовж осей  $a$ і $b$.
Теоретичні результати для  $g$ факторів, що базуються на вище
згаданій локальній структурі, непогано узгоджуються з
експериментальними даними.
\keywords кристалічні поля і гамільтоніани, електронний
парамагнітний резонанс, $\text{Cu}^{2+}$,
$\text{PrBa}_{2}\text{Cu}_{3}\text{O}_{6+x}$
\end{abstract}

\end{document}